\documentclass{jfm}

\usepackage{graphicx}
\usepackage{epstopdf, epsfig}
\usepackage{graphicx}
\usepackage{dcolumn}
\usepackage{bm}
\usepackage{color}
\usepackage{array}
\usepackage{amsmath,amssymb,stmaryrd} 
\usepackage{multicol}

\definecolor{blue}{rgb}{0, 0.4470, 0.7410}
\definecolor{red}{rgb}{0.8500, 0.1250, 0.0480} 
\definecolor{orange}{rgb}{0.8500, 0.3250, 0.0980} 
\definecolor{yellow}{rgb}{0.9290, 0.6940, 0.1250}
\definecolor{purple}{rgb}{0.4940, 0.1840, 0.5560}
\definecolor{green}{rgb}{0.4660, 0.6740, 0.1880}
\definecolor{ltblue}{rgb}{0.3010, 0.7450, 0.9330}
\definecolor{dkred}{rgb}{0.6350, 0.0780, 0.1840}
\definecolor{gray}{rgb}{0.22, 0.22, 0.3}

\newcommand{\bs}{\boldsymbol}

\shorttitle{Phase-response analysis of synchronization for periodic flows}
\shortauthor{K. Taira \& H. Nakao}

\title{Phase-response analysis of synchronization for periodic flows}

\author{Kunihiko Taira\aff{1}\corresp{\email{ktaira@fsu.edu}} \and Hiroya Nakao\aff{2}}

\affiliation{
\aff{1} Department of Mechanical Engineering, Florida State University, Tallahassee, FL 32310, USA
\aff{2} Department of Systems and Control Engineering, Tokyo Institute of Technology, Tokyo 152-8552, Japan
}

\begin{document}

\maketitle

\begin{abstract}
We apply the phase-reduction analysis to examine synchronization properties of periodic fluid flows.  The dynamics of unsteady flows are described in terms of the phase dynamics reducing the high-dimensional fluid flow to its single scalar phase variable.  We characterize the phase response to impulse perturbations, which can in turn quantify the influence of periodic perturbations on the unsteady flow.  These insights from the phase-based analysis uncover the condition for synchronization.  In the present work, we study as an example the influence of periodic external forcing on unsteady cylinder wake. The condition for synchronization is identified and agrees closely with results from direct numerical simulations.  Moreover, the analysis reveals the optimal forcing direction for synchronization.  The phase-response analysis holds potential to uncover lock-on characteristics for a range of periodic flows. 
\end{abstract}


\begin{keywords}
synchronization, phase dynamics, wakes
\end{keywords}


\section{Introduction}

Synchronization is a phenomenon where a system with oscillatory dynamics exhibits its oscillations in unison \citep{Winfree:JTB67, Guckenheimer:JMB75, Kuramoto84, Strogatz:PhysD00, Pikovsky01}.  Such phenomenon was first reported by Huygens, the inventor of the pendulum clock, who noticed that pendulums of two different clocks on a beam always swung in phase or antiphase \citep{Pikovsky01}.  Since then, there have been a variety of studies on uncovering the synchronization mechanisms of oscillators, including oscillatory chemical reactions \citep{Kuramoto84}, synchronous flashing of fireflies \citep{Buck:QRB88}, and jet lag through inter-neuronal communications \citep{Yamaguchi:Science13}.
 
Even if the overall physics involving synchronization is complex and high-dimensional, the analysis of synchronization can be simplified immensely for weak perturbations by considering its phase dynamics.  Such formulation allows us to focus solely on the scalar phase of the oscillator instead of having to be concerned of the full state dynamics.  This approach is known as the phase-reduction analysis and enables an elegant formulation to examine a set of interacting oscillators \citep{Winfree:JTB67, Ermentrout10, Kuramoto84, Nakao:ContempPhys16}.  The phase-based analysis reveals its strength in analyzing networked dynamics and the emergence of synchronized motion and patterns.

While the phase-reduction analysis has been applied to numerous problems in biological \citep{Winfree:JTB67, Ermentrout10} and chemical systems \citep{Kuramoto84}, there has been limited use in the field of unsteady fluid dynamics, especially with regard to synchronization.  Some of the few uses of phase-based analysis to fluid mechanics have been concerned with Hele--Shaw flows \citep{Kawamura:Chaos13, Kawamura:PhysicaD15}.  In the present work, we apply the phase-based analysis to full nonlinear periodic wake flows to examine synchronization.  The discussion herein provides an alternative perspective on examining fluid flows through a single scalar phase variable, which enables significant simplification to the theoretical analysis and reduction in computational and experimental efforts to examine lock on.  

In what follows, we present the phase-reduction analysis \citep{Nakao:ContempPhys16} in the context of periodic fluid flows.  The perspective to examine fluid flows with phase dynamics is provided in section \ref{sec:phase_analysis}, with focus on analyzing synchronization of fluid flows to external excitation.  The phase-reduction analysis is then applied to an example of unsteady cylinder wake with periodic forcing in section \ref{sec:cylinder}.  The condition for synchronization is determined from phase-response analysis revealing the Arnold tongue for lock on.  Concluding remarks are offered in section \ref{sec:conclusion}.  The approach described below can be performed through simulations and experiments, making it attractive for analyzing a range of fluid flow problems, including unsteady flows, active flow control, fluid-structure interactions, and reacting flows.


\section{Phase-reduction analysis}
\label{sec:phase_analysis}

\subsection{Phase-reduction analysis of periodic flows}

\begin{figure}
   \centering
   \includegraphics[width=0.88\textwidth]{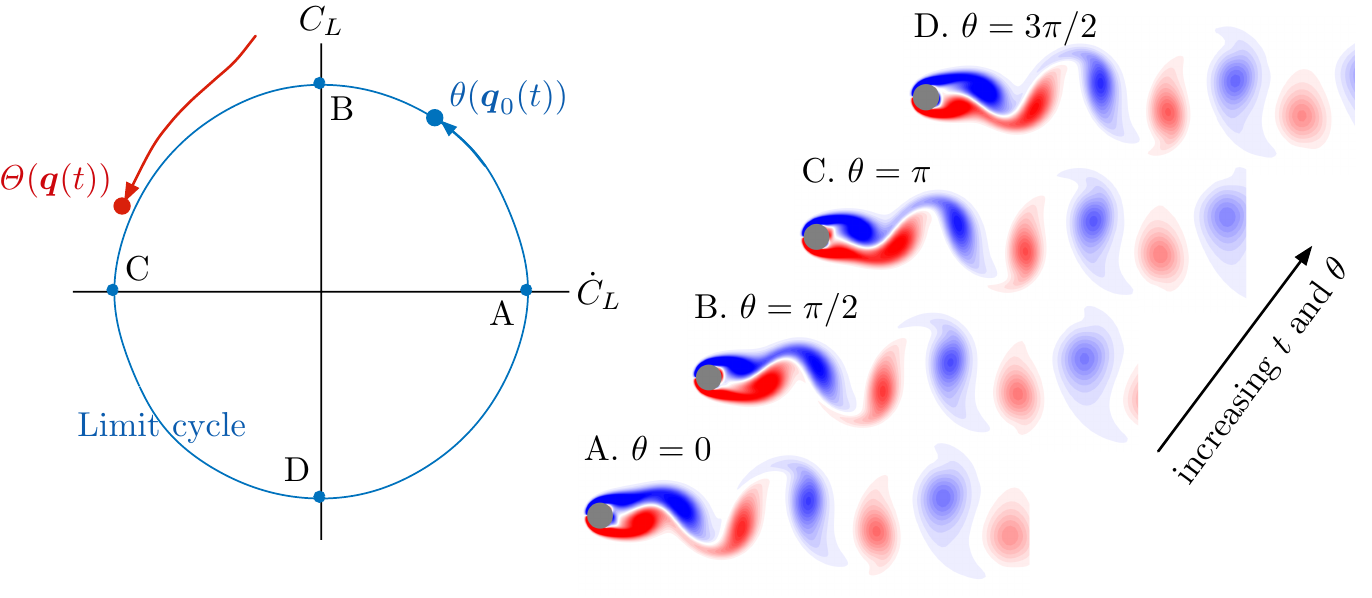}
   \caption{Definition of phase $\theta$ and phase function $\Theta$ with $C_L$ and $\dot{C}_L$ for cylinder flow at $Re = 100$.  Vorticity fields are shown at representative phases.}
   \label{fig:phase_def}
\end{figure}

Let us consider the dynamics of incompressible flow represented as
\begin{equation}
   \dot{\bs{q}} = \bs{F}(\bs{q}),
   \label{eq:gov}
\end{equation}
where $\bs{q}$ is the flow variable that has a stable periodic flow (limit cycle) $\bs{q}_0$ satisfying
\begin{equation}
   \bs{q}_0(\bs{x},t+T) = \bs{q}_0(\bs{x},t).
   \label{eq:periodic}
\end{equation}
For this limit cycle, the periodicity is denoted by $T$ that gives the natural frequency of $\omega_n = 2\pi/T$.  As expressed in equation (\ref{eq:periodic}), flows are generally described in terms of spatial and temporal variables.  For incompressible flow, we can take $\bs{q}$ to be the velocity vector.  In this work, we instead describe the state of the oscillatory flow using the phase $\theta$ of the limit cycle such that
\begin{equation}
   \dot{\theta}(t) = \omega_n,
   \quad \text{where} \quad
   \theta(t) \in [-\pi, \pi].
   \label{eq:freq}
\end{equation}
The main idea here is to reduce the description of the flow from the flow state $\bs{q}(\bs{x},t)$ to a single scalar phase variable $\theta$ \citep{Winfree:JTB67, Guckenheimer:JMB75, Kuramoto84}.  If we have a periodic flow, the phase can point back to the full periodic flow field $\bs{q}_0(\theta)$.  The description of the phase is given in figure \ref{fig:phase_def} for an example of periodic cylinder flow, which will be examined in detail later in section \ref{sec:cylinder}.  In this example, the phase dynamics of the flow can be understood by examining the lift $C_L$ on the cylinder, which holds the same synchronization properties as that of the flow.  The limit cycle of periodic cylinder flow is shown in blue over the $\dot{C}_L$--$C_L$ phase space, with the vorticity fields visualized at representative phases.  

The above characterization of the phase is limited to flow states that reside on the limited cycle $\bs{q}_0$.  We can extend the concept of phase in the basin of the limit cycle and define a phase function $\Theta(\bs{q}(\bs{x},t))$ that returns the phase, i.e., $\theta = \Theta(\bs{q})$.  The phase function can capture the phase dynamics off the limit cycle as illustrated in red in figure \ref{fig:phase_def}.  By combining equations (\ref{eq:gov}) and (\ref{eq:freq}), we observe that the phase function satisfies
\begin{equation}
   \dot{\theta}(t) = \dot{\Theta}(\bs{q}) = \nabla_{\bs{q}} \Theta(\bs{q}) \cdot \dot{\bs{q}} = \nabla_{\bs{q}} \Theta(\bs{q}) \cdot \bs{F}(\bs{q}) = \omega_n.
   \label{eq:phase_fnc}
\end{equation}
This phase function enables the phase to be defined over the entire phase space.  A surface of constant phase is referred to as an isochron and is directed perpendicular to $\nabla \Theta(\bs{q})$ \citep{Guckenheimer:JMB75}.   The notion of isochron is closely related to the eigenfunction of the Koopman operator \citep{Mauroy:Chaos12}.


\subsection{Phase-reduction analysis of perturbed flows}

Now, let us consider perturbing the governing dynamics by
\begin{equation}
   \dot{\bs{q}} = \bs{F} (\bs{q}) + \varepsilon \bs{f}(t), 
\end{equation}
where we have added a small perturbation $\varepsilon \bs{f} (t)$ with $\varepsilon \ll 1$ and $\| \bs{f} \| = 1$.  The dynamics of the phase under perturbation can be determined through
\begin{equation}
   \dot\theta (t)
   = \dot\Theta(\bs{q}(t))
   = \nabla_{\bs q} \Theta({\bs q}) \cdot \dot{{\bs q}}(t)
   = \nabla_{\bs q} \Theta({\bs q}) \cdot \left[ \bs{F} (\bs{q}) + \varepsilon \bs{f}(t) \right].
\label{eq:phase_reduction}
\end{equation}
We can evaluate $\nabla_{\bs q} \Theta$ along the limit cycle with $\bs{q}_0$, assuming that the higher-order terms are negligible since the added perturbation is weak.  Recalling equation (\ref{eq:phase_fnc}), we then arrive at
\begin{equation}
   \dot\theta (t) = \omega_n + \varepsilon {\bs Z}(\theta) \cdot {\bs f}(t),
   \label{eq:phase_forced}
\end{equation}
where 
$ {\bs Z}(\theta) \equiv \nabla_{\bs{q}} \left.\Theta({\bs q}) \right|_{{\bs q} = {\bs q}_0}$ is called the phase-sensitivity function.  Given the perturbation function, we can determine the influence of the perturbation function on the phase dynamics through equation (\ref{eq:phase_forced}) if $\bs{Z}(\theta)$ is known.  

This phase-sensitivity function can be determined in two ways \citep{Nakao:ContempPhys16}.  We can find ${\bs Z}(\theta)$ by examining the phase response to impulse perturbations (direct method) or considering the adjoint dynamics (adjoint method; \cite{Ermentrout10}).  In this work, we consider the former impulse response approach that can be used in both computational and experimental studies.  We consider here the influence of a weak impulse perturbation ${\bs I} = I \delta(t-t_0) h (\boldsymbol{x}) \hat{\bs{e}}_f$ on the state variable $\bs{q}$, where $\hat{\bs{e}}_f$ is the unit vector in the forcing direction and $h(\boldsymbol{x})$ describes the spatial profile of the impulse.  This impulse can be added to the governing dynamics such that  
\begin{equation}
  \dot{\bs q} = {\bs F}({\bs q}) + {\bs I}
\end{equation}
at some phase $\theta$ (at a chosen $t_0$).  With the introduction of this impulse, the phase dynamics becomes affected causing a shift in the phase along the limit cycle state at steady oscillating conditions.  

\begin{figure}
   \centering
   \includegraphics[width=0.86\textwidth]{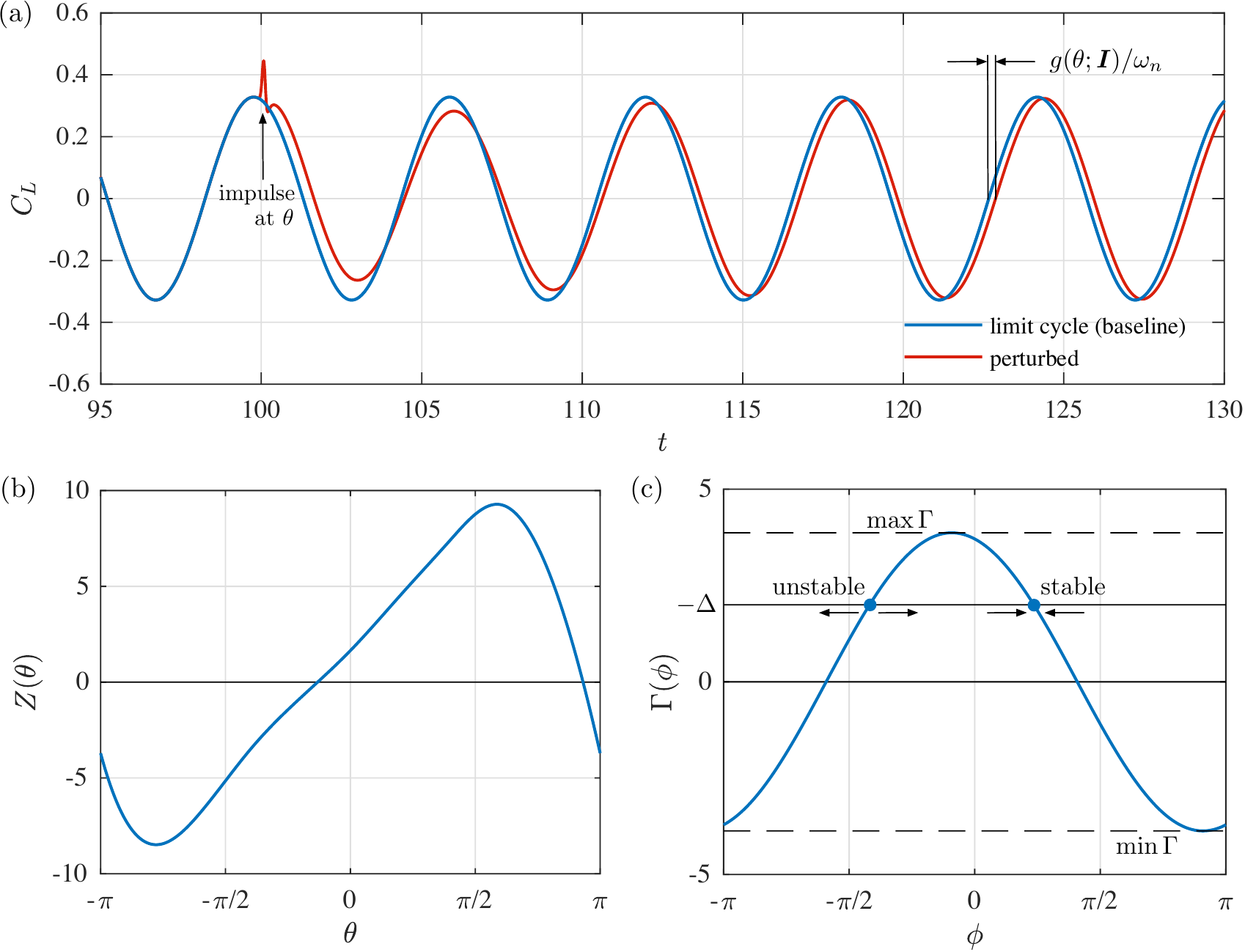}
   \caption{(a) Determining the phase-response function $g(\theta; \bs{I})$ from impulse input.  (b) Phase-sensitivity function.  (c) Phase-coupling function and stability condition for phase difference $\phi$ between the system dynamics and oscillatory actuation.  For graphical clarity, $I=0.025$ is used here.  All subsequent phase-reduction analysis uses $I = 0.01$.}
   \label{fig:phase_analysis}
\end{figure}

The phase change at asymptotic state in response to the impulse is known as the phase-response function $g(\theta; I \hat{\bs{e}}_f)$ and is dependent on the phase at which the impulse is added as well as its location and strength, as illustrated in figure \ref{fig:phase_analysis}(a).  Given a weak perturbation, the phase-sensitivity function can be evaluated as 
\begin{equation}
   Z_j(\theta) = \lim_{I\rightarrow 0} \frac{g(\theta; I \hat{\bs{e}}_f)}{I}
   \approx \frac{g(\theta; I \hat{\bs{e}}_f)}{I}.
   \label{eq:PSF}
\end{equation}
By introducing the impulse perturbations over a range of $\theta$, the phase-sensitivity function can be determined, as shown in figure \ref{fig:phase_analysis}(b).  If the phase dynamics of the state variable $\boldsymbol{q}$ is followed by an observable (sensor output), the observable such as lift in figure \ref{fig:phase_analysis}(a) can be used to determine the phase response.  This enables experimental efforts with limited measurements determine the phase-sensitivity function without difficulty.

\subsection{Synchronization of phase dynamics to external oscillation}

With the oscillatory dynamics of fluid flow reduced to its phase dynamics, we can further analyze the condition for which the flow exhibits synchronization to an external periodic perturbation $\bs{f}(t)$ with a frequency $\omega_f$ (and period $T_f = 2\pi/\omega_f$) that is close to $\omega_n$.  To facilitate our analysis, we consider the relative phase $\phi(t)$ between the phase $\theta(t)$ of the flow and the phase of the perturbation $\omega_f t$
\begin{equation}
   \phi(t) \equiv \theta(t) - \omega_f t.
\end{equation}
Using equation (\ref{eq:phase_forced}), we find the time rate of change of the relative phase is
\begin{equation}
   \dot{\phi}(t)
   = \omega_n - \omega_f + \varepsilon {\bs Z}(\phi(t)+\omega_f t) \cdot {\bs f}(t) 
   = \varepsilon \left[ \Delta + {\bs Z}(\phi(t)+\omega_f t) \cdot {\bs f}(t) \right],
 \label{eq:phasediff}
\end{equation}
where the difference in the natural frequency $\omega_n$ and the perturbation frequency $\omega_f$ is expressed as $ \varepsilon \Delta = \omega_n-\omega_f$ with $\Delta = \mathcal{O}(1)$.

By examining the stability characteristics, we can determine when the oscillatory dynamics of the flow can synchronize to external oscillation.  That is, when $| \dot{\phi} | \rightarrow 0$, the oscillation frequency of the flow converges towards the actuator frequency.  At this moment, equation (\ref{eq:phasediff}) describes the dynamics of the phase difference $\phi(t)$ in a non-autonomous manner.  To perform a stability analysis, it is desirable to have equation (\ref{eq:phasediff}) expressed in autonomous form.  By realizing that $\phi(t)$ varies slowly compared to $\omega_f t$, we can approximate the right-hand side of equation (\ref{eq:phasediff}) by its average over a period of actuation \citep{Kuramoto84, Ermentrout10} to arrive at
\begin{equation}
    \dot{\phi}(t) = \varepsilon \left[ \Delta + \Gamma(\phi) \right],
    \label{eq:phasediff_dyn}
\end{equation}
where 
\begin{equation}
    \Gamma(\phi) 
    = \frac{1}{T_f} \int_t^{t+T_f} \bs{Z}(\phi+\omega_f \tau) 
    \cdot \bs{f}(\tau){\rm d}\tau
    = \frac{1}{2\pi} \int_{-\pi}^{\pi} \bs{Z}(\phi+\psi) 
    \cdot \bs{f}(\psi/\omega_f){\rm d}\psi.
    \label{eq:PCF}
\end{equation}
The function $\Gamma(\phi)$ is $2\pi$-periodic and is known as the phase-coupling function.  With equation (\ref{eq:phasediff_dyn}), we now have a one-dimensional autonomous evolution equation for $\phi$.  

Through inspection of equation (\ref{eq:phasediff_dyn}), we find that the dynamics exhibits stable behavior if $\text{min}_{\phi} \Gamma(\phi) < - \Delta < \text{max}_{\phi} \Gamma(\phi)$, which translates to
\begin{equation}
   \varepsilon \, \text{min}_{\phi} \Gamma(\phi)
   < \omega_f-\omega_n < 
   \varepsilon \, \text{max}_{\phi} \Gamma(\phi).
   \label{eq:sync}
\end{equation}
The phase-coupling function $\Gamma(\phi)$ is shown for the example of cylinder flow in figure \ref{fig:phase_analysis}(c) with an illustration of the stability condition.  The value of the right hand side of equation (\ref{eq:phasediff_dyn}) describes the directions of the solution with respect to the equilibrium points, which are indicated by the arrows in figure \ref{fig:phase_analysis}(c).  These conditions determine the region of stability over the $\omega_f$--$\varepsilon$ space, known as the Arnold tongue \citep{Arnold97}.  This will be shown later in Section \ref{sec:results} (see figure \ref{fig:arnold}).  The decay of frequency difference between the natural limit cycle and the forcing input constitutes the synchronization of the dynamics (they can have a constant offset in phase under synchronization).  Strictly speaking, the above condition for synchronization holds for small forcing input $\varepsilon \ll 1$.  While the above analysis assumes that $\omega_f$ and $\omega_n$ to be close to one another for the averaging operation to be a valid approximation, the analysis can be easily extended to consider subharmonic and superharmonic frequencies.

For fluid flows, such synchronization amounts to the so called lock on.  The above result enables us to determine the combination of actuation frequency and amplitude for the flow to synchronize with periodic excitations.  As long as we can examine the system response against impulse inputs, the overall formulation can be performed through numerical simulations or experiments, which is attractive for fluid mechanics.


\section{Phase-response analysis of unsteady cylinder wake}
\label{sec:cylinder}

\subsection{Model problem}

\begin{figure}
   \centering
   \includegraphics[width=0.88\textwidth]{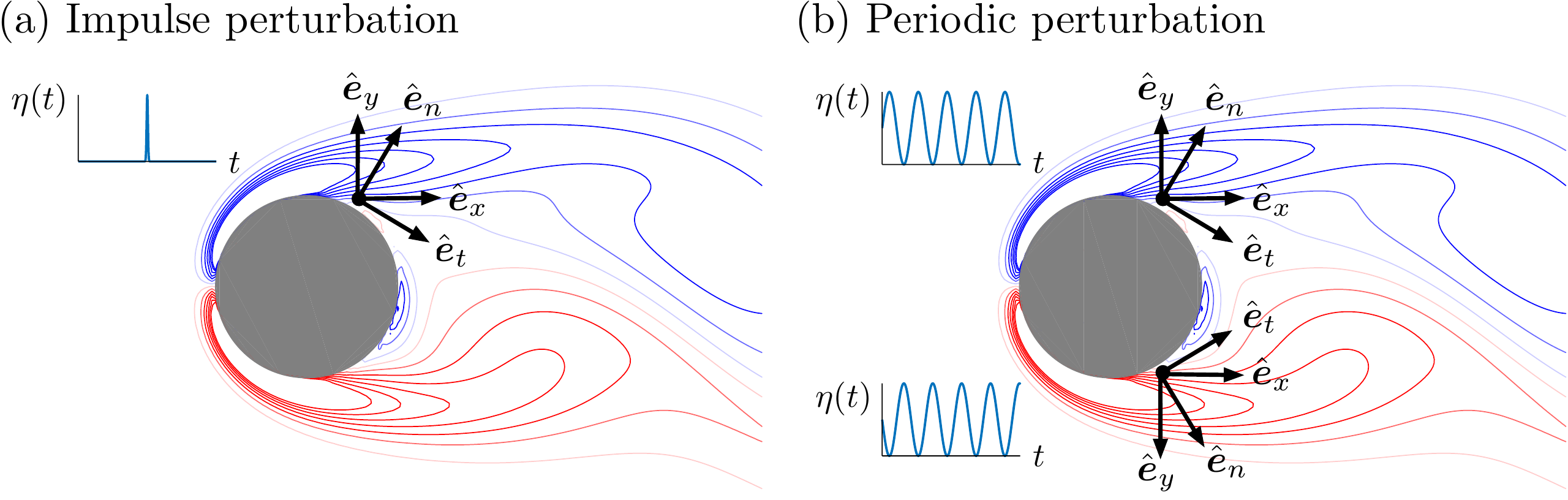}
   \caption{Cylinder flow forcing setup to analyze (a) the impulse response with top actuator only and (b) the effect of periodic forcing with top and bottom actuators.  Selected forcing directions are $x$ direction ($x$), $y$ direction ($y$), tangential ($t$), and normal ($n$).}
   \label{fig:control_setup}
\end{figure}

We apply the phase-reduction analysis to examine the synchronization of circular cylinder wake to periodic forcing.  The base periodic flow is taken to be the two-dimensional incompressible flow over a circular cylinder at a diameter-based Reynolds number of $Re \equiv U_\infty d/\nu = 100$, where $U_\infty$, $d$, and $\nu$ denote the characteristic freestream velocity, cylinder diameter, and kinematic viscosity of the flow, respectively.  The dynamics of the flow (equation (\ref{eq:gov})) is governed by the non-dimensional incompressible Navier--Stokes equations:
\begin{equation}
    \frac{\partial \bs{u}}{\partial t} + \bs{u}\cdot\nabla{\bs u} 
    = - \nabla p + Re^{-1} \nabla^2 {\bs u} + \varepsilon \bs{f}
    \quad
    \text{and}
    \quad
    \nabla \cdot {\bs u} = 0,
\end{equation}
where $\bs{u}$ is the velocity and $p$ is the pressure.  In this equation, the forcing (actuation) input $\varepsilon \bs{f}$ is taken to be a localized force (momentum injection)
\begin{equation}
    \varepsilon \bs{f} 
    =  \delta(\bs{x}-\bs{x}_f) \eta(t) \hat{\bs e}_f
\end{equation}
having a magnitude of $\varepsilon \ll 1$, position at $\bs{x}_f$, and forcing direction specified by a unit vector $\hat{\bs{e}}_f$.  The Dirac delta in space is approximated using the three-cell discrete delta function \citep{Roma:JCP99}.   The overall setups of the forcing input for the two cases considered below are illustrated in figure \ref{fig:control_setup}.  

The function $\eta(t)$ specifies the temporal form of the forcing input as
\begin{equation}
   \eta(t) = 
   \begin{cases} 
   I \delta(t) \approx \frac{I}{\sqrt{2\pi}\sigma} \exp\left[- \frac{1}{2}(\frac{t - t_0}{\sigma})^2\right]   	
   & \text{for impulse perturbation} \\
   \frac{\varepsilon}{2} \left[ 1 + \sin(\omega_f t + \psi) \right] 			
   & \text{for periodic perturbation}
   \end{cases} 
\end{equation}
The impulse perturbation, approximated as a narrow Gaussian in time (with $\sigma = 0.005$), is introduced to the flow to determine the phase-response, phase-sensitivity, and phase-coupling functions.  The sinusoidal forcing function is chosen to examine the lock-on behavior of the flow to periodic flow control input.  While the impulse perturbation is introduced only at the top separation point ($58.4^\circ$ from the aft separation point, positioned $3 \Delta x$ away from the surface), we add the sinusoidal forcing input at the top and bottom separation points with a phase difference of $\phi$ (top: $\psi = 0$ and bottom: $\psi = \pi$).  The actuation is introduced near the separation points to effectively modify the wake dynamics.  However, the perturbations can be added to any locations in the flow field for this analysis.  The forcing amplitudes are chosen such that $I = 0.01$ and $\varepsilon \in [0, 0.1]$.  The corresponding steady and oscillatory momentum coefficients for the periodically forced cases are less than $0.02$ and $0.007$, respectively, for $\varepsilon \le 0.1$ \citep{Munday:PF13}.


\subsection{Computational Setup}

The immersed boundary projection method simulates the unsteady flow on a Cartesian grid with the cylinder represented through the boundary forcing \citep{Taira:JCP07}.  For the present study, the fast version of the immersed boundary projection method is chosen \citep{Colonius:CMAME08} with the computational domain extending over $(x/d, y/d) \in [-31,33] \times [-32,32]$.  The grid near the cylinder has $\Delta x_\text{min}/d = \Delta y_\text{min}/d = 0.02$ and the time step is chosen to meet $U_\infty \Delta t/\Delta x_\text{min} = 0.5$.  The computational algorithm and setup have been validated extensively in past studies, including a similar study on control of cylinder flow \citep{Munday:PF13}.  The present study captures the baseline flow accurately and agrees with findings from three independent numerical codes, as summarized in Table \ref{table:validation}.

\begin{table}
\centering
   \begin{tabular}{llll}
   					& $C_L$ 		& $C_D$			& $St$		\\ 
   Present				& $\pm 0.328$	& $1.345\pm 0.009$	& $0.165$		\\ 
   \cite{Liu:JCP98} 		& $\pm 0.339$ 	& $1.35\pm 0.012$ 	& $0.165$ 	\\
   \cite{Linnick:JCP05}	& $\pm 0.337$	& $1.34\pm 0.009$	& $0.165$ 	\\ 
   \cite{Canuto:JFM15}	& $\pm 0.329$	& $1.34\pm 0.0091$	& $0.167$ 	\\ 
   \end{tabular}
   \caption{Comparison of lift, drag, and Strouhal number for circular cylinder flow at $Re = 100$.}
   \label{table:validation}
\end{table}


\subsection{Analysis of cylinder wake}
\label{sec:results}

In this section, we determine the phase-response, phase-sensitivity, and phase-coupling functions to characterize the cylinder flow dynamics in terms of phase on the $\dot{C}_L$--$C_L$ space.  Once these functions are determined, we can find the conditions for synchronization of the cylinder wake to periodic perturbations.  

We first characterize the phase response of cylinder wake to weak impulse perturbations added near the top separation point.  By evaluating the change in phase, we can determine the phase-response function $g(\theta; I \hat{\bs{e}}_f)$.  The magnitude of the impulse is set to $I = 0.01$, which is small in magnitude compared to that of the force on the cylinder.  Here, we consider adding the impulse in four different directions ($x$, $y$, surface-tangential, and surface-normal directions) from the top separation point, as illustrated in figure \ref{fig:control_setup}.  These directions are chosen to form two pairs of orthogonal directions, such that they can serve as bases for subsequent analyses.  

\begin{figure}
\centering
   \includegraphics[width=0.96\textwidth]{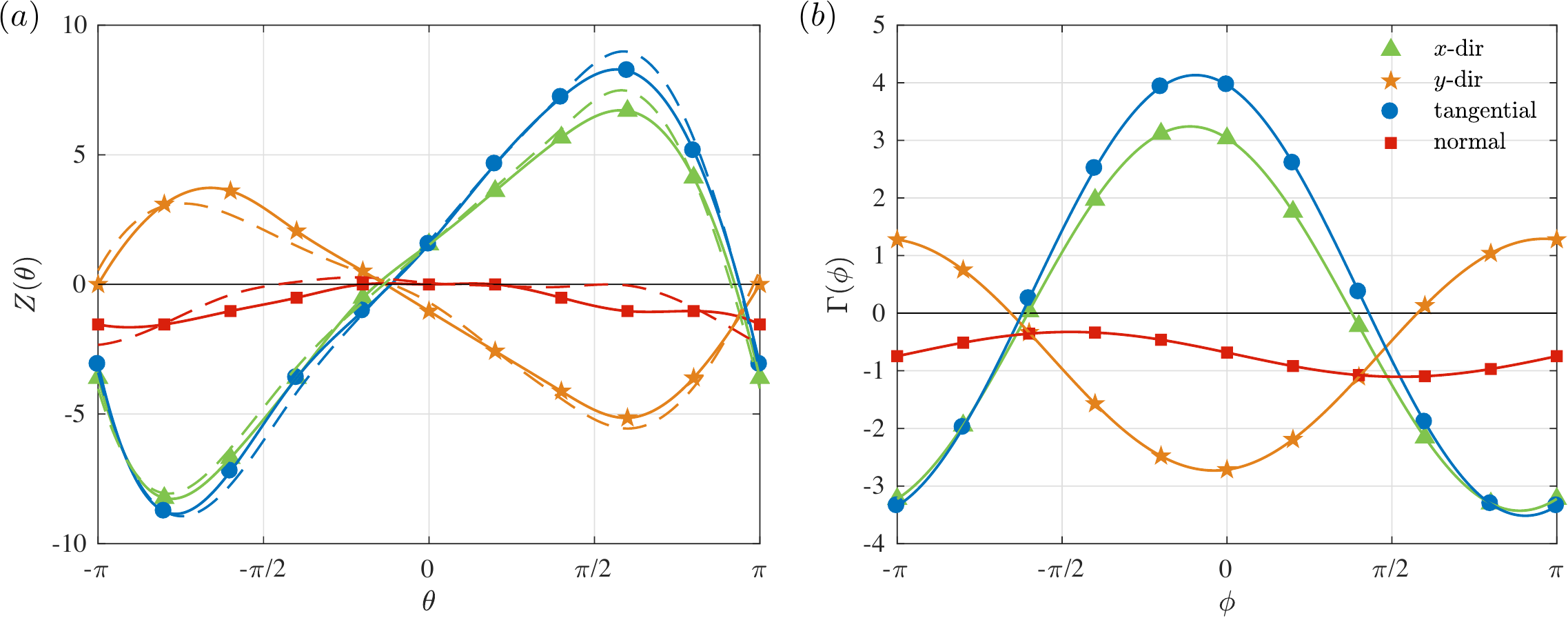}
   \caption{(a) Phase-sensitivity function $Z(\theta)$ and (b) phase-coupling function $\Gamma(\phi)$ for cylinder wake and periodic forcing input.  The dashed lines for the phase-sensitivity function show approximations from complementary bases.  Symbols represent where actual simulations were performed.}
   \label{fig:PSF}
\end{figure}

Based on the impulse responses, we can compile in figure \ref{fig:PSF}(a) the phase-sensitivity function $Z(\theta)$ evaluated with equation (\ref{eq:PSF}) over the phase space for the four directions, shown by the solid lines.  Symbols represent where actual simulations were performed; 10 simulations were performed for each direction.  Impulse added in the tangential and $x$ directions lead to phase advancement.  The tangential forcing exhibits the largest phase sensitivity among the four directions examined.  Forcing in the $y$ direction causes phase delay and normal forcing does not result in much of a phase change.  We also show with dashed lines in figure \ref{fig:PSF}(a) the phase-sensitivity functions found by rotating the coordinate systems.  That is, $x$ and $y$ components are evaluated based on tangential and normal components, and vice versa.  The close agreements between the solid and dashed lines suggest that the chosen impulse magnitude is adequately small for the linear approximations made in equations (\ref{eq:phase_forced}) and (\ref{eq:PSF}) to hold.  Hence, it suffices to perform the impulse-response analysis in two independent forcing directions to map out the phase-sensitivity function over the phase space.  In this example, the phase sensitivity function quantifies the sensitivity of the shear layer dynamics aft of the separation point (as well as the overall wake vortex dynamics) to the added impulse perturbation.

With the phase-sensitivity function $Z(\theta)$ available, the phase-coupling function $\Gamma(\phi)$ can be found through equation (\ref{eq:PCF}).  For the present periodic forcing case, we need to sum the influence of the top and bottom actuators, which essentially doubles the magnitude of $\eta(t)$ in the computation of $\Gamma(\phi)$.  The phase-coupling functions for the four representative directions are shown in figure \ref{fig:PSF}(b).  As expected, large variations in $\Gamma(\phi)$ over the relative phase $\phi$ are given by the periodic perturbations in the tangential and $x$ directions.  Forcing in the $y$ direction has opposing effects with reduced variation.  Normal forcing seems to be the least influential among these four directions.  It is interesting to note that $Z(\theta)$ has a zero average over $\theta$ for the tangential forcing case, which removes the influence of non-zero mean of the forcing input.

\begin{figure}
   \centering
   \includegraphics[width=0.96\textwidth]{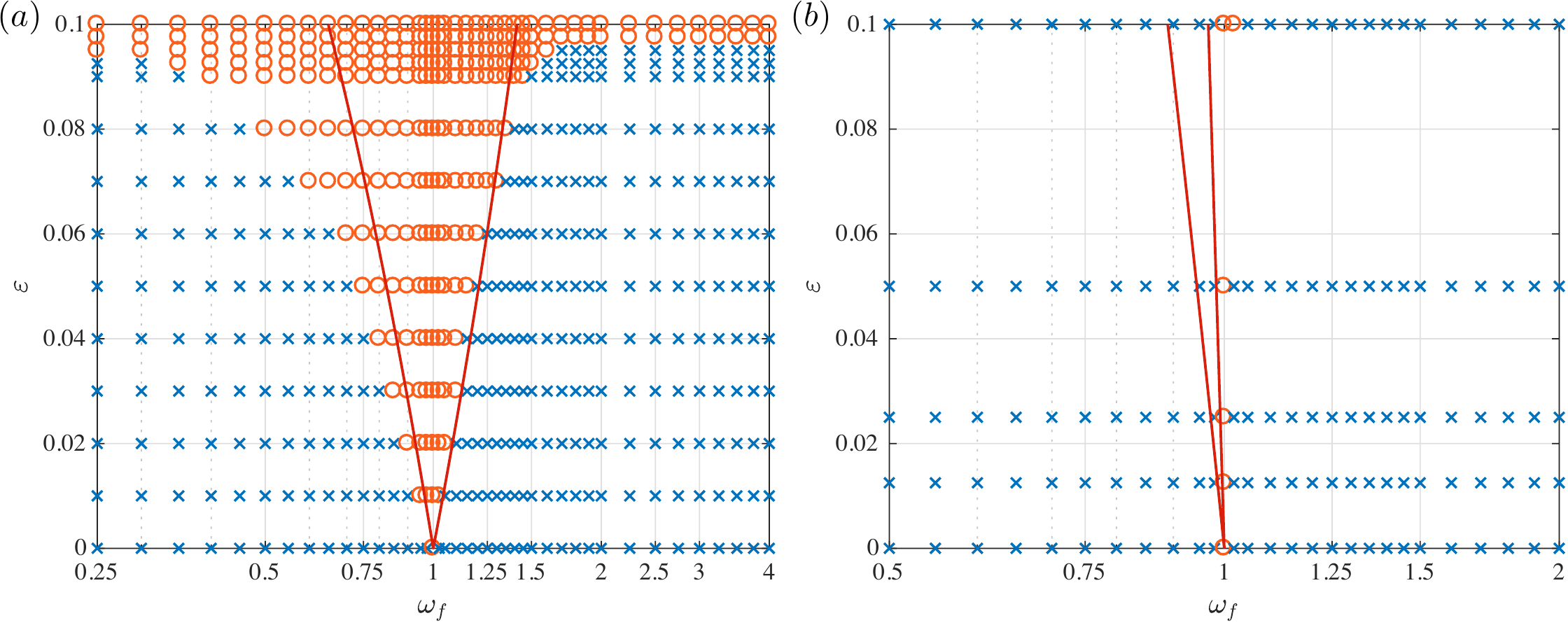}
   \caption{Synchronization of cylinder wake to periodic actuation for (a) tangential and (b) normal forcing inputs.  Synchronized and unsynchronized cases from DNS \citep{Munday:PF13} are shown by $\circ$ and $\times$, respectively.  The criteria for synchronization (\ref{eq:sync}) correspond to the region bounded by the red lines.}
   \label{fig:arnold}
\end{figure}

Recall that greater variations in the phase-coupling function widens the regions for achieving synchronization of the wake dynamics to actuation fluctuation, as stated in condition (\ref{eq:sync}).  For example, tangential and normal direction forcing yield $(\min \Gamma, \max \Gamma) = (-3.52,4.13)$ and $(-1.10, 0.33)$, respectively.  We can take these minimum and maximum values for the synchronization condition and compare them with the results from DNS \citep{Munday:PF13}, as summarized in figure \ref{fig:arnold}.  In the work of \cite{Munday:PF13}, the cylinder wake is considered to be synchronized (locked on) to the actuation frequency when the lift power spectra exhibits a single peak at the actuation frequency.  

For the tangential forcing cases shown in figure \ref{fig:arnold}(a), the condition for synchronization from the phase-response analysis agrees well with the full nonlinear wake simulations capturing the region of the Arnold tongue, especially for low forcing amplitude $\varepsilon$ until nonlinear effects become apparent at higher $\varepsilon \gtrsim 0.07$.  As mentioned earlier, the phase-sensitivity function for tangential forcing removes the influence of the non-zero forcing input for synchronization.  In fact, the region of lock on is not altered much for a zero mean forcing case, as presented in the DNS findings \citep{Munday:PF13}.  For $\varepsilon \gtrsim 0.097$, the wake becomes synchronized to the oscillatory forcing for all considered actuation frequencies, which is beyond the validity of the present phase-reduction analysis.  

For the normal forcing case, we observe that the Arnold tongue is much narrower than that of the tangential forcing case, as evident from figure \ref{fig:arnold}(b).  The phase-reduction analysis predicts the narrow width of synchronization region well, but slightly biases the tongue towards the lower frequency, which becomes obvious for higher forcing amplitudes.  
This deviating trend from normal forcing is likely caused by the inaccuracy in the
evaluation of the small-amplitude phase-sensitivity function as well as by
the nonlinear effects.

\begin{figure}
   \centering
   \includegraphics[width=0.95\textwidth]{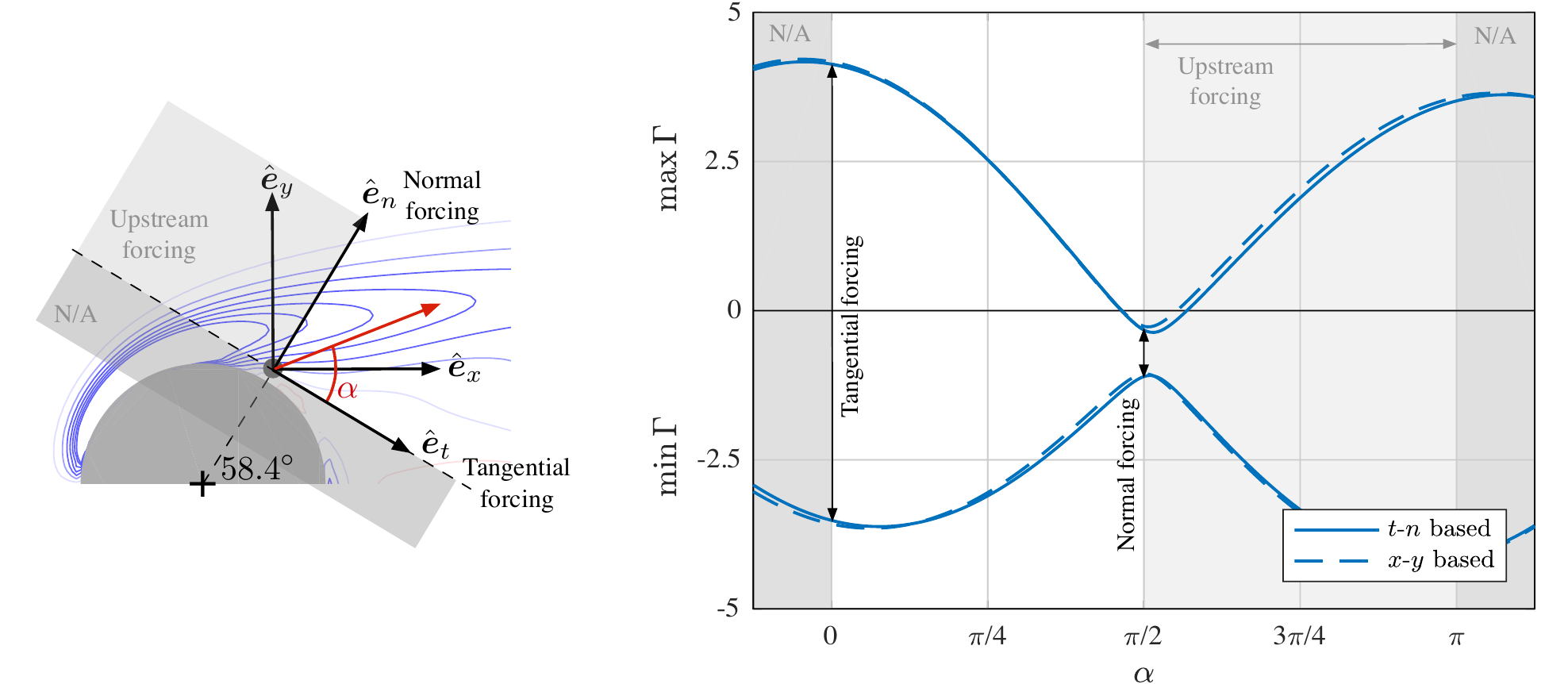}
   \caption{Minimum and maximum values of the phase-coupling function $\Gamma$ over the forcing angle $\alpha$ measured the tangential direction.  Extrema based on the bases of $x$ and $y$ directions as well as tangential and normal directions are shown.  Forcing directions and the angle $\alpha$ are depicted on the left.}
   \label{fig:minmaxGamma}
\end{figure}

The identification of the Arnold tongue from the phase-response analysis is attractive since we only require a limited number of simulations to determine the phase-response function over the phase $\theta \in [-\pi, \pi]$.  Instead of having to perform a large number of full DNS (or experiments) as a parametric study to locate the Arnold tongue, we can analytically determine the boundary of synchronization from the phase-coupling function.  Moreover, for two-dimensional flow, we can consider two independent forcing directions to determine the impulse response for any forcing direction, as depicted in figure \ref{fig:minmaxGamma}.  By constructing the phase-sensitivity function over the forcing direction $\alpha$ based on the tangential and normal components or the $x$ and $y$ components, we can determine the minimum and maximum values of $\Gamma$ as a function of $\alpha$, as shown in figure \ref{fig:minmaxGamma}.  In this diagram, we shade out the directions that would be directed below the tangential direction (into the body) as well as the upstream direction (acting against the free stream).  Restricting our interest to downstream forcing between $\alpha \in [0, \pi/2]$, we find that synchronization can be achieved for the widest range of $\omega_f$ with tangential forcing ($\alpha \approx 2^\circ$).  On the other hand, normal forcing would be the most difficult to achieve synchronization with its narrowest bound over $\alpha$.  We emphasize that the bounds for synchronization has been determined with only a limited number of simulations and does not require an extensive parameter sweep, which significantly reduces the need for computational (and experimental) resources.

While synchronization in this particular example leads to drag increase for the majority of the synchronized cases, the phase-response analysis shows tremendous potential to identify the condition of lock on for a range of fluid flow problems.   In this example, we know from our previous work \citep{Munday:PF13} that the addition of non-zero mean in tangential forcing can achieve drag reduction and with minimal effect of altering the lock-on region.  In fact, the zero-mean case also shows good agreement with the predicted condition for synchronization.  There are other cylinder flow examples that can achieve drag reduction, as in the case of synchronizing the wake to sinusoidal rotational of the cylinder \citep{Tokumaru:JFM91}.

For the presently considered periodically forced cylinder wake, we do not observe subharmonic or superharmonic synchronizations from DNS.  There are of course other fluid flows that exhibit such behavior.  Phase-response analysis can be extended to determine the condition for synchronization near harmonic frequencies.


\section{Conclusion}
\label{sec:conclusion}

We have applied the phase-reduction analysis to periodic fluid flows for studying the phase dynamics and synchronization properties.  This was achieved by capturing the oscillatory physics through its scalar phase dynamics.  With such formulation, we were able to determine the phase response to impulse perturbations, which formed the basis to evaluate the influence of external perturbations on the oscillatory flow dynamics.  These insights reveal the Arnold tongue over which the flow can exhibit synchronization to periodic excitation.  As an example, we have considered the unsteady wake dynamics behind a circular cylinder and characterized the synchronization properties for momentum-based forcing.  Comparison with DNS showed close agreement with the phase-based analysis.  The present study also confirmed that the tangential forcing direction is the optimal direction to achieve lock on for the widest range of actuation frequency.  This phase-based analysis can be performed in numerical simulations and experiments, holding promising potential to uncover synchronization properties for a large class of problems, including wake dynamics, flow control, vortex-induced vibration, fluid-structure interaction, and reacting flows, without the need for extensive parametric sweeps.  


\section*{Acknowledgments}

KT gratefully acknowledge the support from the US Air Force Office of Scientific Research (Grant: FA9550-18-1-0040, Program Manager: Dr.~Douglas R.~Smith) and the US Army Research Office (Grant: W911NF-17-1-0118, Program Manager: Dr.~Matthew J.~Munson).  HN acknowledges financial support from the Japan Society for the Promotion of Science KAKENHI Grants: JP16H01538, JP16K13847, and JP17H03279. 


\bibliographystyle{jfm}
\bibliography{Taira_refs}

\end{document}